\documentclass[journal]{IEEEtran}
\pdfoutput=1

\usepackage[utf8]{inputenc}
\usepackage[english]{babel}
\usepackage[pdftex]{graphicx}
\DeclareGraphicsExtensions{.pdf,.jpeg,.png}
\usepackage{amsmath}
\usepackage{array}
\ifCLASSOPTIONcompsoc
  \usepackage[caption=false,font=normalsize,labelfont=sf,textfont=sf, subrefformat=parens,labelformat=parens]{subfig}
\else
  \usepackage[caption=false,font=footnotesize, subrefformat=parens,labelformat=parens]{subfig}
\fi

\usepackage[nolist]{acronym}
\usepackage[binary-units=true, detect-weight=true, detect-family=true]{siunitx}
\usepackage{booktabs}
	
\usepackage[switch, modulo]{lineno}

\usepackage{scalerel}
\usepackage{tikz}
\usetikzlibrary{svg.path}

\definecolor{orcidlogocol}{HTML}{A6CE39}
\tikzset{
  orcidlogo/.pic={
    \fill[orcidlogocol] svg{M256,128c0,70.7-57.3,128-128,128C57.3,256,0,198.7,0,128C0,57.3,57.3,0,128,0C198.7,0,256,57.3,256,128z};
    \fill[white] svg{M86.3,186.2H70.9V79.1h15.4v48.4V186.2z}
                 svg{M108.9,79.1h41.6c39.6,0,57,28.3,57,53.6c0,27.5-21.5,53.6-56.8,53.6h-41.8V79.1z M124.3,172.4h24.5c34.9,0,42.9-26.5,42.9-39.7c0-21.5-13.7-39.7-43.7-39.7h-23.7V172.4z}
                 svg{M88.7,56.8c0,5.5-4.5,10.1-10.1,10.1c-5.6,0-10.1-4.6-10.1-10.1c0-5.6,4.5-10.1,10.1-10.1C84.2,46.7,88.7,51.3,88.7,56.8z};
  }
}

\newcommand{\orcidicon}[1]{\href{https://orcid.org/#1}{\mbox{\scalerel*{
\begin{tikzpicture}[yscale=-1,transform shape]
\pic{orcidlogo};
\end{tikzpicture}
}{|}}}}

\usepackage[backend=biber,bibencoding=utf-8,style=ieee,sorting=none,doi=true,isbn=false,url=false,minbibnames=1,maxbibnames=1,mincitenames=1,maxcitenames=1]{biblatex}
\usepackage[babel]{csquotes}
\usepackage{hyperref}
\uspunctuation
\bibliography{./misc/literature.bib}

\begin{document}
%
\title{A novel DOI Positioning Algorithm for Monolithic Scintillator Crystals in PET based on Gradient Tree Boosting}


\author{{\center\IEEEauthorblockN{Florian Müller \orcidicon{0000-0002-9496-4710}\ , David Schug \orcidicon{0000-0002-5154-8303}\ , Patrick Hallen \orcidicon{0000-0001-9371-0025}\ , Jan Grahe, Volkmar Schulz \orcidicon{0000-0003-1341-9356}}}%
\thanks{Manuscript submitted August 15, 2018; revised October 6, 2018; accepted November 20, 2018. (Corresponding author: Florian Müller.)}
\thanks{This project is funded by the German Research Foundation (DFG), project number SCHU 2973/2-1.}
\thanks{The authors are with the Department of Physics of Molecular
Imaging Systems, Institute for Experimental Molecular Imaging,
RWTH Aachen University, 52074 Aachen, Germany (e-mail:
florian.mueller@pmi.rwth-aachen.de).}
\thanks{Digital Object Identifier 10.1109/TRPMS.2018.2884320}}

\markboth{IEEE Transactions on Radiation and Plasma Medical Sciences}%
{Shell \MakeLowercase{\textit{et al.}}: Bare Demo of IEEEtran.cls for IEEE Transactions on Magnetics Journals}
%



\IEEEtitleabstractindextext{%
\begin{abstract}
Monolithic crystals are examined as an alternative to segmented scintillator arrays in positron emission tomography (PET). Monoliths provide good energy, timing and spatial resolution including intrinsic depth of interaction (DOI) encoding. DOI allows reducing parallax errors (radial astigmatism) at off-center positions within a PET ring. We present a novel DOI-estimation approach based on the supervised machine learning algorithm gradient tree boosting (GTB). GTB builds predictive regression models based on sequential binary comparisons (decision trees). GTB models have been shown to be implementable in FPGA if the memory requirement fits the available resources. We propose two optimization scenarios for the best possible positioning performance: One restricting the available memory to enable a future FPGA implementation and one without any restrictions. The positioning performance of the GTB models is compared with a DOI estimation method based on a single DOI observable (SO) comparable to other methods presented in literature. For a 12 mm high monolith, we achieve an averaged spatial resolution of 2.15 mm and 2.12 mm FWHM for SO and GTB models, respectively. In contrast to SO models, GTB models show a nearly uniform positioning performance over the whole crystal depth.
\end{abstract}

}

\maketitle

\IEEEdisplaynontitleabstractindextext

%
\IEEEpeerreviewmaketitle

\section{Introduction}

\IEEEPARstart{P}{ET} is a functional imaging technique with a variety of applications in both preclinical as well as clinical research and practice \cite{Phelps2004, Myers2001}. Two \SI{511}{\kilo\eV} gamma particles originating from a positron-electron annihilation are registered by radiation detectors arranged in a ring geometry. State-of-the-art radiation detectors consist of scintillation crystals (e.g., BGO, LSO, LYSO) converting the gamma particles to optical photons and photosensor arrays with multiple channels detecting the optical photons.

The key challenge in PET detector instrumentation is to detect the gamma particles with high sensitivity and with good spatial, energy and time resolution. The knowledge of \acused{DOI}\ac{DOI} reduces the parallax error (radial astigmatism) at off-center positions within a PET ring \cite{Saha2016}. \ac{DOI} information is especially important for PET systems with a small ring diameter such as preclinical devices or organ-specific applications imaging the brain or female breast \cite{Gonzalez2016a, Lee2017, Krishnamoorthy2018}. Recently, the advantages of \ac{DOI} information have also been experimentally validated for a clinical case with a \SI{70}{\centi\metre} diameter tomographic setup \cite{Borghi2018}. Furthermore, \ac{DOI} is expected to also improve the performance of long-axial field-of-view PET systems (Total Body PET) \cite{Zhang2018}. 

Several detector concepts have been demonstrated in literature which can be mainly divided up into two groups: pixelated and monolithic detectors. Most state-of-the-art clinical PET scanners are operated with pixelated detector designs. These detectors consist of arrays of scintillator needles and are read out employing light-sharing techniques or one-to-one coupling. None of the currently used clinical detectors for whole-body PET provide \ac{DOI} information \cite{Vandenberghe2016}. Several adaptations of pixelated detectors have been presented to obtain \ac{DOI} information. Among others, multiple-layer designs with small shifts between each layer \cite{Ito2010}, light-sharing techniques encoding the \ac{DOI} information into the light distribution \cite{Lee2015}, combination of scintillation layers with different properties (phoswich detectors) \cite{Seidel1999}, sub-surface laser engraving \cite{Uchida2016} and dual-sided readout \cite{Kang2015} were proposed. However, often these concepts come at the cost of other key performance parameters or introduce additional complexity and cost to the detector design.

As an alternative, monolithic detectors consist of scintillators without any segmentation coupled directly to a photosensor array. Monoliths are able to provide good spatial, timing and energy resolutions as widely shown in literature \cite{Borghi2016,VanDam2013,Bruyndonckx2008,marcinkowski2016sub,Gonzalez2016}. Furthermore, monolithic detectors enable intrinsic DOI encoding. For example, the three-dimensional position can be estimation based on a fit of a model of the light distribution \cite{Li2008}. Instead of fitting to a model, maximum likelihood searches based on three-dimensional training data have also been successfully demonstrated \cite{Krishnamoorthy2018, Marcinkowski2016}. A different approach employed an analytic expression based on the attenuation of the scintillator and the expected light-transport \cite{Gonzalez-Montoro2018, Gonzalez-Montoro2017}. After fitting two free parameters, the ratio of event energy to the maximum local intensity was utilized as DOI observable. A method based on a single \ac{DOI} observable only including the photon counts - the sum of the squared pixel intensities - was applied to both single and dual-sided read-out \cite{Borghi2016, Borghi2016a}. The expected distribution of events according to the attenuation was simulated and then matched to the distribution of the \ac{DOI} observable. The main principle of this method testing several \ac{DOI} observables was first published in \cite{vanDam.2011_practicalMethod}. Besides the presented methods, also neural network estimators based on training data acquired by a side irradiation of the scintillator were shown \cite{Wang2013}. To widely translate monolithic scintillators into applications with a large number of detectors, time-efficient and easy calibration methods are required. Furthermore, all employed algorithms such as position estimation need to be scalable for a large number of detectors.

Recently, we presented a novel planar positioning algorithm based on the supervised machine learning technique \ac{GTB} \cite{Muller2018a}. Here, we present the adaptation and application of this algorithm to \ac{DOI} calibration by a side irradiation of the scintillator. \ac{GTB} builds a set of sequential binary decisions (decision trees) which are evaluated as simple comparisons with two possible outcomes. The algorithm handles different sets of input features and their combinations as well as partially missing data. Trained \ac{GTB} models are shown to be implementable in FPGA if the memory requirement does not exceed the available resources \cite{Kuaga, VanEssen2012}. Besides a general description of the hyperparameter tuning of \ac{GTB} models, we present two optimization scenarios to find the best possible positioning performance: one restricting the available memory to enable a future FPGA implementation and one without any restrictions. The positioning performance is compared to a DOI-approach using a single \ac{DOI} observable similar to \cite{Borghi2016} as well as results presented in literature.

\section{Materials}

As the same materials and detector were used as presented in \cite{Muller2018a}, only a brief description of the single components is given in the corresponding sections. We utilized the \ac{PDPC} technology evaluation kit (TEK) with two sensor tiles of DPC 3200-22 photon counter as a coincidence setup. For calibration, a fan beam collimator providing adjustable beam widths for both coincidence and detector under study was employed. The fan beam collimator consists of a fixed bottom shielding, a tool housing up to two sources, and two adjustable top shielding units defining the beam (see Fig. \ref{fig:sketch_setup}). The whole setup was operated in a light-tight temperature chamber at approx. \SI{5}{\celsius} sensor temperature.

\subsection{Photodetector}

We used an array, referred to as tile, of \num{4x4} digital Silicon Photomultiplier (dSiPM) DPC 3200-22 of \ac{PDPC} \cite{Degenhardt2009, Frach2009}. Each DPC is an independent trigger region consisting of \num{4} pixels with a pitch of \SI{4}{\milli\metre} resulting in a total of \num{64} pixels per tile. Every DPC provides a customizable two-level trigger scheme: After the first trigger signal is generated, the second, higher threshold needs the be reached within a given time interval as well. For our settings, on average \num{2.33} and \num{17} photons need to be detected to reach the first and second trigger threshold. In case the second trigger condition is met, the integration phase starts and all \num{4} pixels of the DPC are read out afterward. The collected information of a DPC is referred to as a hit. The entity of all hits corresponding to one gamma particle interaction is called a cluster as described in more detail in Sec. \ref{subsec:data_acqusition}. For a single gamma particle interaction, not all \num{16} DPCs of the tile will certainly output hit data. This leads to clusters with missing hits as reported in Sec. \ref{sec:results_data_acquisition}, especially if the photon densities are low. Further information regarding the photosensor can be found in \cite{Schug2012, Tabacchini2014, Marcinkowski2013, Schaart2016}.

\subsection{Scintillator Crystal and Wrapping}

We studied a monolithic LYSO scintillator of dimensions \SI{32x32x12}{\milli\metre} matching the active sensor area of the tile. The monolith was wrapped in highly reflective Teflon\texttrademark{} tape (Klinger, Idstein, Germany) and coupled to the photosensor using a two-component dielectric silicon gel (Sylgard 527, Dow Corning, Midland, Michigan, USA). We chose the reflective wrapping to achieve a high light-output of the scintillator. An optical simulation studying the influence of several scintillator wrappings (e.g., black tape) on the position performance of \ac{GTB}-models can be found in \cite{Grahe2017a}. To register coincidences, we employed a \SI{12}{\milli\metre}-high pixelated array with \SI{1}{\milli\metre} pitch also utilized in \cite{Schug2015, Gross-Weege2016}.

\subsection{Collimator Setup}

\begin{figure}[t!]	
\center
\includegraphics[width=2.7in]{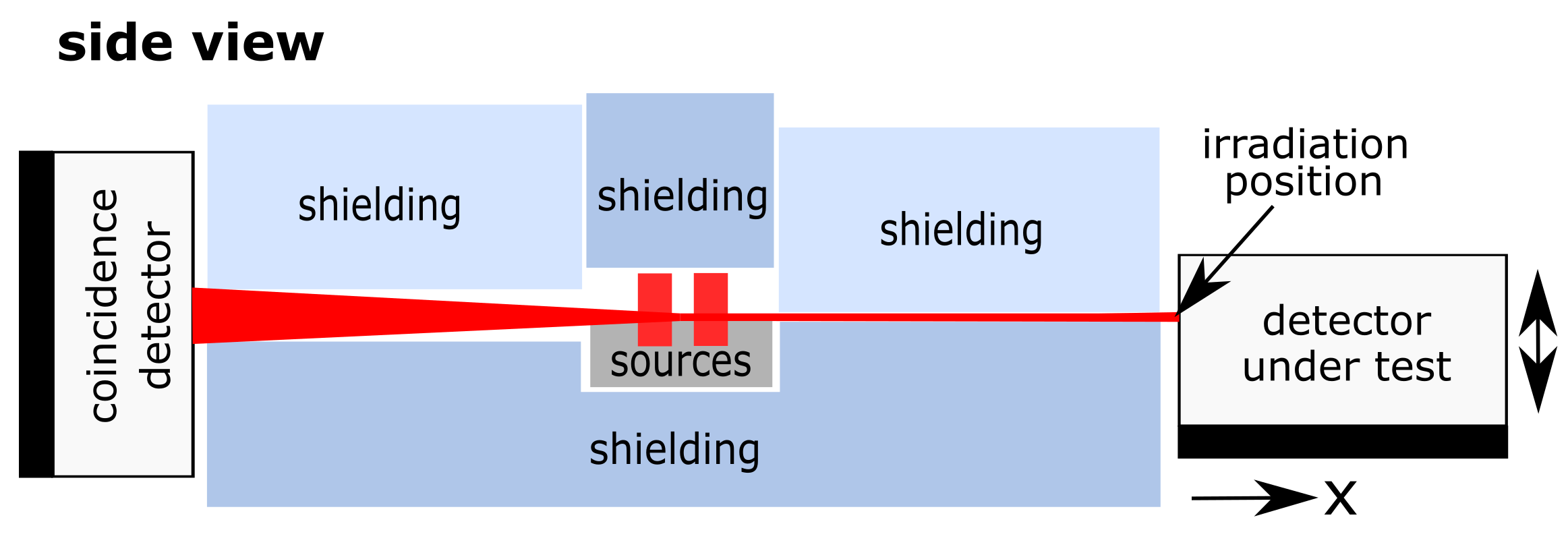}
\caption{Sketch of the setup containing the fan beam collimator, coincidence detector and detector under test. For \ac{DOI} calibration, a side irradiation of the detector under test is performed. The irradiation position is determined by the feed-back loop of the stepper motor.}
\label{fig:sketch_setup}
\end{figure}

The setup contained a fan beam collimator equipped with two $^{22}\text{Na}$ sources of approximately \SI{10}{\mega\becquerel} each and an electrically driven translation stage (LIMES 90, Owis, Staufen im Breisgau, Germany) (see Fig. \ref{fig:sketch_setup}). All measurements were conducted at a sensor temperature of approx. \SI{5}{\celsius} achievable on system level as demonstrated in \cite{weissler2015hyperionIID}. The fan beam collimator reached a coincidence rate of \SI{199}{\hertz} for slit widths of \SI{5}{\milli\metre} and \SI{0.25}{\milli\metre} for the coincidence detector and the detector under study, respectively. We chose a much larger slit width for the coincidence detector to avoid losing coincidence events due to geometrical effects and uncertainties of the collimator setup. The slit width of \SI{0.25}{\milli\metre} for the detector under study translates to a beam width of \SI{0.42}{\milli\metre} FWHM at the crystal surface determined by stepping a crystal edge through the beam \cite{Muller2018a}. The translation stage was connected to a control PC and the current position of the detector under test was recorded. This irradiation position was employed as reference in later analysis.

\section{Methods}

\subsection{Data Acquisition and Preprocessing}
\label{subsec:data_acqusition}

The data acquisition aimed to create datasets for establishing and testing \ac{DOI} positioning models based on a side irradiation of the detector. However, a side irradiation leads to an exponential distribution of the gamma particle interactions along the propagation direction according to Lambert-Beer's law. For the employed setup (see Fig. \ref{fig:sketch_setup}), we expected an exponential distribution along the x-direction and a uniform distribution along the y-direction. To achieve reliable \ac{DOI} positioning models and estimates of the positioning performance, datasets with uniformly distributed events in both x- and y-direction are strongly beneficial. Otherwise, the exponential distribution may mislead machine learning-based algorithms during training of DOI models and weights the positioning performance achieved over the whole crystal volume nonproportional. Therefore, we need a planar positioning algorithm to estimate the gamma interaction position of the events obtained with the side irradiation. Subsequently, the full calibration process included irradiations of the crystal's top surface for planar positioning calibration and a side irradiation for \ac{DOI} positioning calibration. Both x- and y-direction of the crystal were irradiated at parallel lines with a pitch of \SI{0.75}{\milli\metre} for the planar positioning calibration. The side irradiation was performed with a pitch of \SI{0.25}{\milli\metre}.

All conducted measurements shared the initial preprocessing based on a tool developed by Schug et al. \cite{Schug2015}: The collected optical photons of one gamma particle interaction create up to \num{16} hits in each of the detectors. All hits associated to one gamma particle interaction were identified and merged to clusters. We applied a cluster window of \SI{40}{\nano\second}. The timestamp of the earliest hit determined the timestamp of the cluster. Afterward, coincident clusters were searched using a sliding coincidence window of \SI{20}{\nano\second}. Pixels missing in a cluster were marked with a negative value to distinguish them from zero photon counts.

Clusters with a total photon count fewer than \num{700} photons were rejected to remove noisy events from the data. Based on the total photon distribution, this photon cut translates to an energy cut of around \SI{300}{\kilo\eV}. No further quality cuts were applied to the data. Accepted clusters are called events.

In the next step, \ac{GTB}-based planar positioning models were established employing the irradiations of the crystal's top surface as described in \cite{Muller2018a}. All events acquired by the side irradiation were positioned along the planar directions. Then, a subset uniformly distributed along x- and y-direction was chosen for every irradiation position of the side irradiation as motivated above. The resulting data were split up into three data sets: 1) training data for building the \ac{DOI} positioning models; 2) validation data for optimization of the hyperparameter of the \ac{DOI} positioning models; 3) test data for the final evaluation of the positioning performance. Each data set contained \num{10000} events per irradiation position of the side irradiation. For the training data, both the binning of the training data as well as the number of events per irradiation position were varied later on aiming for short calibration times. We selected a minimum number of \num{100} training events per irradiation position due to practical reasons: For \num{100} training events, the required time for moving the translation stage and operations of the control PC are already the dominating factor compared to an irradiation time of around \SI{0.5}{\second}.

\subsection{Performance Parameters}

All following positioning performance parameters are based on the positioning error distribution (irradiation position - estimated position) calculated for every single irradiation position. If a single value is given, the performance parameter is averaged for all irradiation positions.

\begin{enumerate}
\item Bias Vector: The bias vector is defined as the mean positioning error for a given position. Due to the edge effects found in monolithic crystals, the bias vector magnitude distribution is non-Gaussian. 50th and 90th percentile of the bias vector magnitude distribution ($\text{Bias}_{50}$ and $\text{Bias}_{90}$) are given to account for both the central part and tails of the distribution.
\item \ac{SR}: The \ac{SR} is defined as the FWHM of the positioning error distribution. We calculate the \ac{SR} in accordance to the NEMA NU 4-2008 procedure \cite{Of2010}. The \ac{SR} is not corrected for the finite beam width of the collimator.
\item Root Mean Squared Error (RMSE): The RMSE is the root of the mean squared positioning error.
\item Mean Absolute Error \acused{MAE}(\ac{MAE}): The \ac{MAE} is the mean of the absolute positioning error.
\item Percentile Distance $d_x$: The percentile distance is defined as the distance enclosing the given percentile $x$ of all events around an irradiation position. We report the 50th and 90th percentile distance.
\item Score of distance \SI{1.5}{\milli\metre}: The score of distance \SI{1.5}{\milli\metre} is the fraction of events which are assigned a position within \SI{1.5}{\milli\metre} around the irradiation position.
\end{enumerate}

While \ac{MAE}, RMSE, percentile distances, and the score value are sensitive to bias effects, SR is not prone to a global shift of the positioning error distribution.

\subsection{Single Observable DOI Estimation}

This \ac{DOI} estimation method utilizes a single observable \acused{SO}(\ac{SO}) calculated based on the measured light distribution as demonstrated in \cite{vanDam.2011_practicalMethod, Borghi2016a}. In contrast to methods presented in literature, the observable must not include more than the four pixels of one DPC to avoid jitter caused by missing hits in an event. We defined a set of possible observables and examined the correlation with the \ac{DOI} position. The set included \mbox{1) the} fraction of the highest photon count to the total photon count of the hottest DPC, and 2) the sum of the squared pixel intensities of the hottest DPC. The crystal was divided into equally sized segments with their own calibration to account for differences in the detector response. The number of segments ranged from \numrange{1}{32} along both planar directions.

To match the observable to the \ac{DOI} position, multiple methods such as lookup tables and fits to polynomials of higher order are possible. For this work, we employed \ac{IR}. \ac{IR} minimizes the mean squared error for monotone data without assuming any form of target function \cite{Kakade, Pedregosa2011}. Thus, this method is preferable if no physical model is present.

As this method is a benchmark to compare the \ac{GTB}-based \ac{DOI} estimation models, only results of the best performing observable and segmentation are shown. The averaged performance parameters as well as the spatial distributions of bias vector, MAE and SR are reported. Tests reducing the number of training events and irradiation positions aiming for short calibration times are out of the scope of this paper and are not presented.

\subsection{Gradient Tree Boosting DOI Estimation}

A detailed description of the \ac{GTB} algorithm for planar positioning is given in \cite{Muller2018a}. Thus, only the main characteristics of the algorithm and hyperparameters used later on are described.

As part of supervised machine learning techniques, \ac{GTB} utilizes training data with known irradiation positions to establish predictive regression models. The algorithm handles missing data for training and evaluation and can be used with arbitrary input features \cite{Chen}. \ac{GTB} builds a set of sequential binary decisions (decision trees) which are evaluated as simple comparisons with two possible outcomes \cite{Kotsiantis2013, Natekin2013}. The ensemble is trained in an additive manner: The first decision tree is based on the given irradiation position. Every following decision tree is trained on the positioning error (irradiation position – estimated position) of the previous ensemble. We employed RMSE as training loss of the objective function.

Four hyperparameters of \ac{GTB} models are of particular importance for the following optimizations: 1) Number of decision trees of the ensemble. 2) Maximum depth: The maximum number of comparisons in a single decision tree. 3) Learning rate: The learning rate multiplicates the positioning error of the already established ensemble with a constant factor less or equal \num{1} for training of the next decision tree as described in \cite{Muller2018a}. This allows to reduce the number of decision trees while reducing the highest achievable positioning performance \cite{Friedman2001}. 4) Features of the input set: In addition to the \num{64} raw photon counts, further features motivated by the physical properties of the problem can be added to improve the positioning performance. Used features and input sets are defined in the following section.

Evaluation of trained \ac{GTB} models is possible in CPU-, GPU- and FPGA-based architectures \cite{VanEssen2012}. All listed architectures should allow for real-time event processing. This work focusses on the possibility of an FPGA implementation because FPGAs are widely employed at several points in the current and future architecture developed in this group \cite{weissler2015hyperionIID, GebhardtdigitalFPGAPipeline2012}. FPGA-based processing combined with the data acquisition allows to significantly decrease the amount of data which needs to be sent to the connected server reducing the bandwidth requirements. Trained \ac{GTB} models are implementable in FPGA if the memory requirement does not exceed the available resources \cite{Kuaga, VanEssen2012}. The memory requirement (MR) of a single decision tree of a \ac{GTB} model can be estimated by
\begin{equation}
\text{MR(\textit{d})} = \left(2^d - 1\right)\cdot \SI{11}{\byte} + 2^d \cdot\SI{6}{\byte}
\end{equation}
with $d$ the maximum depth \cite{Muller2018a}.
 
We present a general optimization protocol examining the influence of the single hyperparameters. Furthermore, two optimization scenarios of GTB models for the best possible positioning performance are demonstrated: One restricting the available memory to enable a future FPGA implementation and one without any restrictions.

\subsubsection{General Optimization Process}

We adapted the developed optimization protocol presented in \cite{Muller2018a} to the \ac{DOI} problem. First, a suitable start point was searched testing similar settings found for the planar positioning as in \cite{Muller2018a}. Aiming for short calibration times, we studied the influence of the binning of the irradiation positions as well as of the number of training events per irradiation position. The binning of the irradiation positions ranged from \SIrange{0.25}{3}{\milli\metre} and the number of events per irradiation position from \numrange{100}{10000}. Then, always one of the hyperparameters introduced in the previous section was varied keeping all other parameters constant. We tested maximum depths from \numrange{4}{12}, learning rates from \numrange{0.05}{0.7} and three different input sets defined below. The parameter ranges were chosen according to the results of the planar positioning optimization and suggestions found in literature \cite{Muller2018a, Natekin2013}. In general, \ac{GTB} models were trained for \num{1000} decision trees and tested with the validation data set. 

To study the influence of the input sets, we validated three possible sets of input features: 1) Raw data: the 64 photon counts; 2) Raw data and calculated features (CF): The calculated features included the index numbers of the hottest pixel and DPC, first and second moment of the light distribution, both defined \ac{DOI} observables, the total photon sum and projections of the photon counts along both planar directions; 3) Raw data, CF and estimated planar interactions positions. In contrast to input set 3), input sets 1) and 2) do not require the estimated planar interaction position for evaluating events. Thus, \ac{GTB} models trained with these input sets can be fully parallelized together with the planar positioning models which could be beneficial for processing data of a full PET system. We present results for the different input sets for all tested maximum depths ranging from \numrange{4}{12}. To represent models with low and high memory requirements and study the influence of the input sets during the training process, ensembles of \num{50} and \num{1000} decision trees were validated.

For all optimization steps, we chose to present the averaged \ac{MAE} as validation metric to account for bias effects and allow a direct comparison with other publications \cite{Borghi2016, Borghi2016a}. Additionally, an overlay and a difference plot of the spatial distribution of the \ac{MAE} for the varied binning of irradiation positions are shown. The spatial distribution indicates if \ac{GTB} models have working regression capabilities: For a working regression model, no bias towards the irradiation positions used for training should be observed.

\subsubsection{High-Performance Optimization}

No memory restrictions were applied to select the best-performing \ac{DOI} positioning models. We elected to pick those models with the best averaged RMSE value for all three tested input sets as the RMSE is used as loss function during model training as well. First, the minimum averaged RMSE was searched for every \ac{GTB} model trained during the general optimization process. In case the minimum RMSE was found for the maximum trained number of decision trees (1000 trees), we continued training until the respective \ac{GTB} model tend to overfit. The \ac{GTB} model was assumed to be overfitting, if the RMSE started to worsen. Second, the best averaged RMSE of all models was searched. The averaged performance parameters of these \ac{GTB} models are presented. Furthermore, the spatial distributions of bias vector, MAE, and SR are plotted for the best-performing model with raw data and calculated features as input.

\subsubsection{Memory-Requirement Optimization}

To enable an FPGA implementation, the \ac{GTB} model with the best possible positioning performance for a given memory restriction is searched. We trained the \ac{GTB} models for combinations of maximum depth and learning rate while an empirically chosen convergence criterion determined the number of decision trees. No further decision trees were added if the averaged MAE did not improve more than \SI{0.0001}{\milli\metre/decision\,tree}. For high learning rates, the chosen convergence criterion may be too conservative and takes action in the overfitting regime. Thus, the found positioning performance is compared to those found for the high-performance optimization. In case the found number of decision trees is larger than those of the high-performance model, the cut gets discarded and the high-performance model of same maximum depth and learning rate is selected.

\section{Results}

\subsection{Data Acquisition}
\label{sec:results_data_acquisition}

The number of read out DPCs per gamma interaction followed a Gaussian distribution with a mean of \num{9.8} DPCs and a standard deviation of \num{1.8} DPCs. An irradiation time per irradiation position of around \SI{5}{\min} was required to measure \num{10000} events uniformly distributed along the planar directions.

\subsection{Single Observable DOI Estimation}
\begin{figure}[!t]
\center
\includegraphics[width=2.5in]{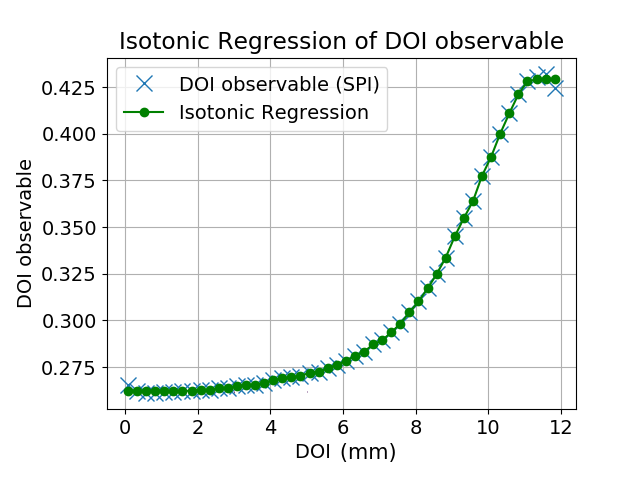}
\caption{Course of the DOI observable squared pixel intensities (SPI) and \ac{IR}. The \ac{DOI} observable monotonely increases with the \ac{DOI} position. \ac{DOI} position \SI{0}{\milli\metre} represents the top surface of the crystal and \SI{12}{\milli\metre} the photosensor.}
\label{fig:comparison_IR_lin_fit}
\end{figure}

\begin{table}[t!]
\center
\caption{Overview of averaged performance parameters for the best found single observable (SO) model and high performance \ac{GTB} models for all three input sets (raw data (r); raw data and calculated features (r+CF); raw data, calculated features and estimated planar interaction positions (r+CF+pos).}
\label{tab:overview_results}
\begin{tabular}{@{}lS[round-mode=places, round-precision=2]S[round-mode=places, round-precision=2]S[round-mode=places, round-precision=2]S[round-mode=places, round-precision=2]@{}}
\toprule
          & {SO}     & \multicolumn{3}{c}{GTB}        \\ \midrule
          &        & {r}        & {r+CF}     & {r+CF+pos} \\\midrule
RMSE / \si{\milli\metre} & 2,2207 & 1,8635   & 1,814625 & 1,809    \\
MAE / \si{\milli\metre}      & 1,7256 & 1,3194   & 1,283739 & 1,277968 \\
SR  / \si{\milli\metre}      & 2,1487 & 2,2584   & 2,120161 & 2,09335  \\
$d_{50}$/ \si{\milli\metre}     & 1,3457 & 0,922421 & 0,844323 & 0,877167 \\
$d_{90}$ / \si{\milli\metre}    & 3,595  & 3,0074   & 2,911626 & 2,90467  \\
$\text{Bias}_{50}$/ \si{\milli\metre} & 1,0559 & 0,540936 & 0,549904 & 0,5474   \\
$\text{Bias}_{90}$ / \si{\milli\metre} & 2,6198 & 1,737724 & 1,672499 & 1,6639   \\
Score of \SI{1.5}{\milli\metre}   & 0,5421 & 0,702886 & 0,714351 & 0,7164   \\ \bottomrule
\end{tabular}
\end{table}

\begin{figure}[t!]
\center
\subfloat[]{\includegraphics[width=2.4in]{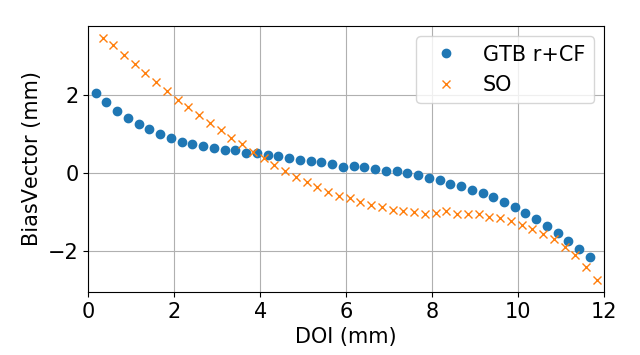}%
\label{fig:result_IR_GTB_bias}}\hfill
\subfloat[]{\includegraphics[width=2.4in]{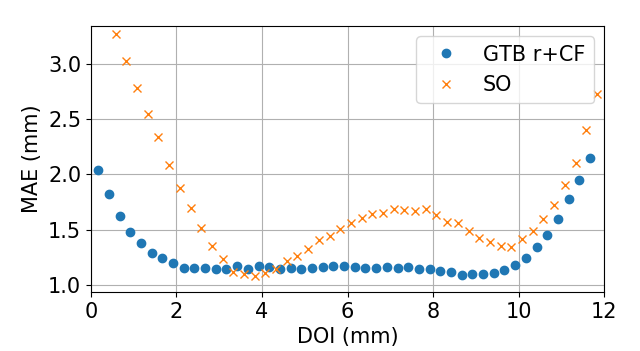}%
\label{fig:result_IR_GTB_MAE}}\hfill
\subfloat[]{\includegraphics[width=2.4in]{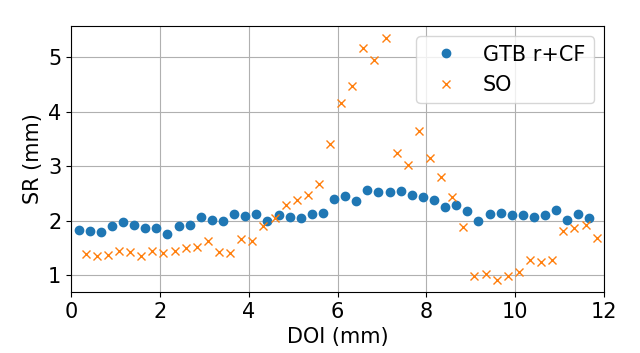}%
\label{fig:result_IR_GTB_SR}}
\caption{Spatial distribution of three performance parameters for the best performing single observable models (denoted as SO) and \ac{GTB} model with raw data and calculated features as input (denoted as GTB r+CF). \ac{DOI} position \SI{0}{\milli\metre} represents the top surface of the crystal and \SI{12}{\milli\metre} the photosensor. (a) Distribution of the bias vector. (b) Distribution of the \ac{MAE}. (c) Distribution of the \ac{SR}.}
\label{fig:result_IR_GTB}
\end{figure}

Both tested \ac{DOI} observables led to similar results. In general, the sum of squared pixel intensities performed a bit better (approx. \SI{3}{\%}). Also, the segmentation along the planar directions improved the positioning performance with an optimum at \num{8} segments along both directions. The difference of positioning performance between the best and worst \ac{DOI} model was less than \SI{2.5}{\%}. Fig. \ref{fig:comparison_IR_lin_fit} shows the course of the chosen \ac{DOI} observable. Assuming a monotone behavior of the \ac{DOI} observable is justified. An averaged \ac{MAE} and \ac{SR} of \SI{1.73}{\milli\metre} and \SI{2.15}{\milli\metre} FWHM were achieved (see Tab. \ref{tab:overview_results}). Fig. \ref{fig:result_IR_GTB} displays the spatial distribution of bias vector, \ac{MAE} and \ac{SR}. Over the whole crystal depth, a bias vector ranging from approx. \SI{\pm 3}{\milli\metre} close to the edges to approx. \SI{1}{\milli\metre} in the central area is observed. The \ac{MAE} deteriorates towards the edges and shows an additional decrease of the positioning performance between \SIrange{4}{10}{\milli\metre} \ac{DOI} position. The \ac{SO} method achieves a \ac{SR} better than \SI{2}{\milli\metre} FWHM for \SIrange{0}{4.5}{\milli\metre} and \SIrange{8.9}{12}{\milli\metre} \ac{DOI} position. However, the positioning performance significantly worsens for the other \ac{DOI} positions to a maximum of \SI{5.4}{\milli\metre} FWHM.

\subsection{Gradient Tree Boosting DOI Estimation}

\subsubsection{General Optimization Process}

\newcommand{\subfigwidth}{2.4in}
\begin{figure*}[!t]
\center
\subfloat[]{\includegraphics[width=\subfigwidth]{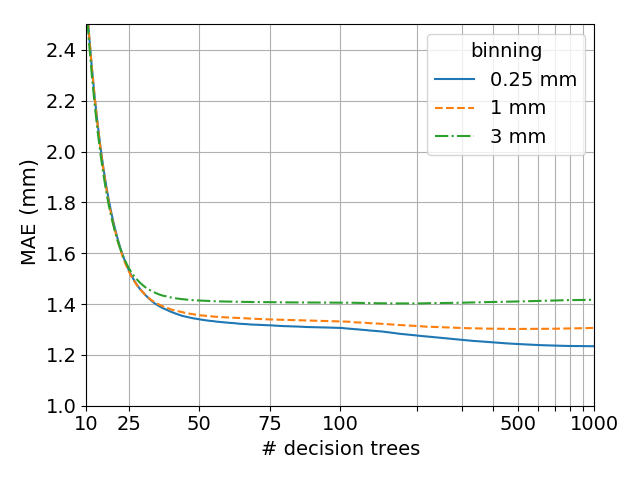}%
\label{fig:genereal_reference_binning}}
\subfloat[]{\includegraphics[width=\subfigwidth]{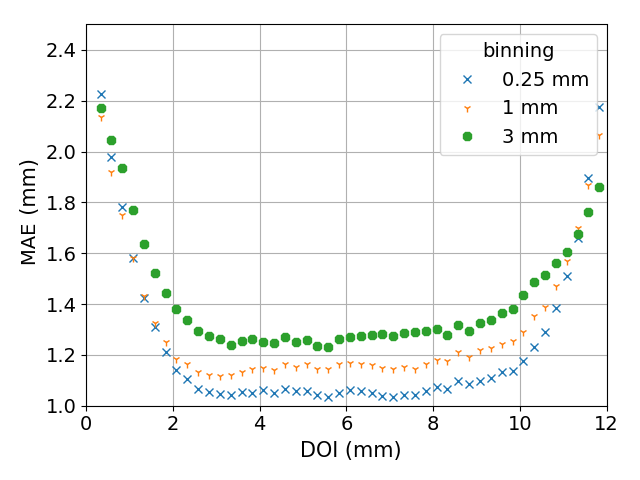}%
\label{fig:general_reference_binning_overlay}}
\subfloat[]{\includegraphics[width=\subfigwidth]{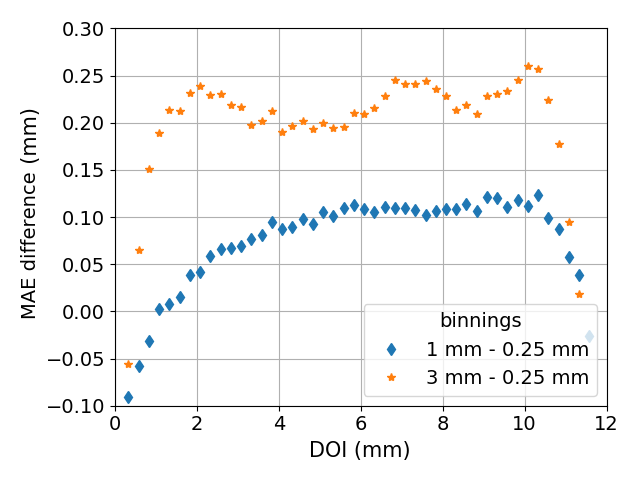}%
\label{fig:general_reference_binning_overlay_diff}}
\hfill
\subfloat[]{\includegraphics[width=\subfigwidth]{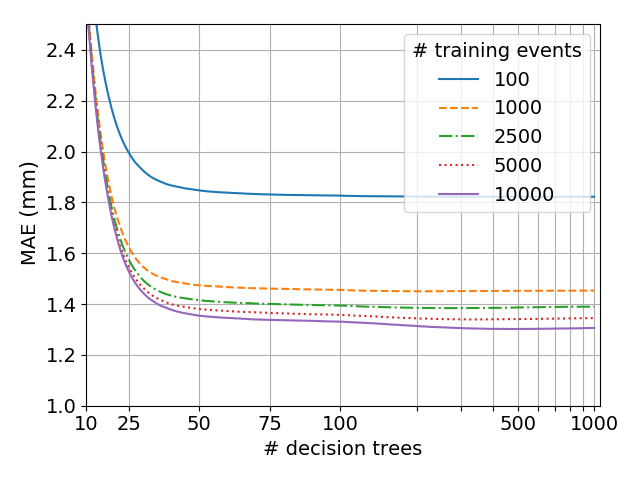}%
\label{fig:general_training_events}}
\subfloat[]{\includegraphics[width=\subfigwidth]{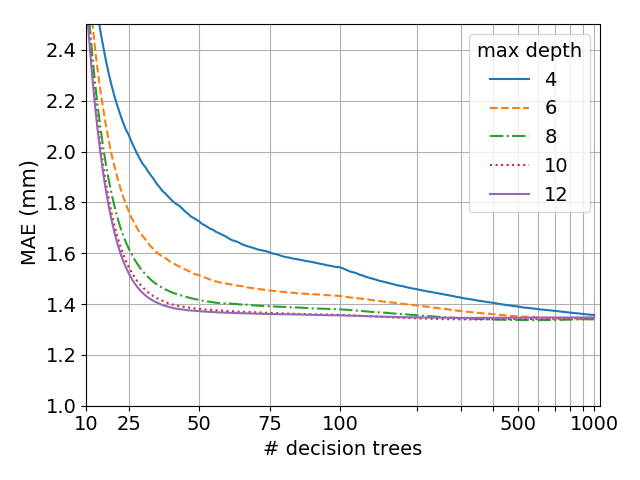}%
\label{fig:general_max_depth}}
\subfloat[]{\includegraphics[width=\subfigwidth]{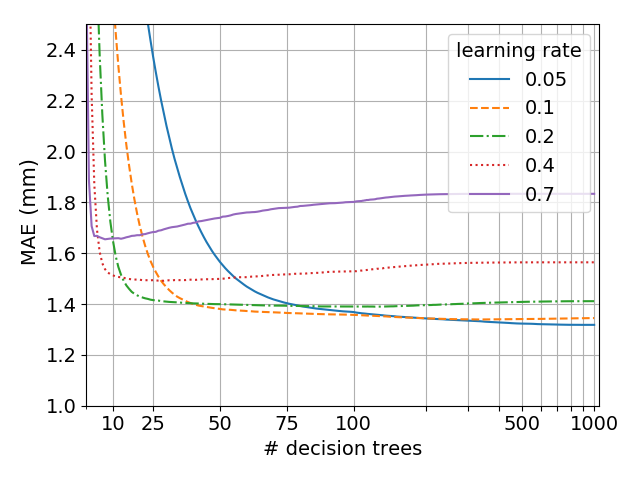}%
\label{fig:general_learning_rate}}
\hfill
\subfloat[]{\includegraphics[width=\subfigwidth]{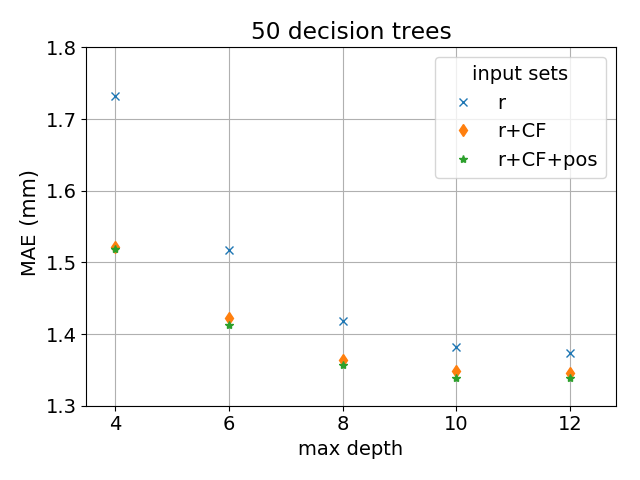}%
\label{fig:general_input_sets_small}}
\subfloat[]{\includegraphics[width=\subfigwidth]{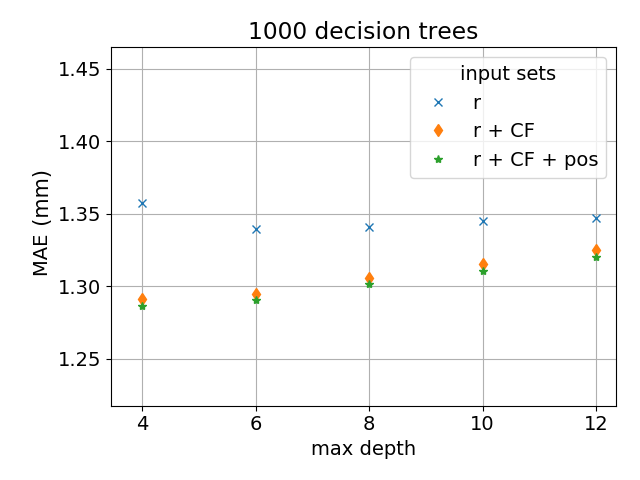}%
\label{fig:general_input_sets_large}}
\caption{General optimization process for \ac{GTB} models. We employed a maximum depth of \num{10}, learning rate \num{0.1}, raw data of \num{5000} training events per irradiation position as input set and pitch of \SI{1}{\milli\metre} of the calibration grid unless stated otherwise. The \ac{MAE} is calculated employing the validation data set with \num{10000} test events and \SI{0.25}{\milli\metre} pitch of the irradiation grid. For the figures which display the number of decision trees, the abscissa is linearly scaled up to \num{100} decision trees and logarithmically afterwards. In case a spatial distribution is shown, \ac{DOI} position \SI{0}{\milli\metre} represents the top surface of the crystal and \SI{12}{\milli\metre} the photosensor. Plots regarding the binning of the calibration grid show only selected pitches of the calibration grid due to reasons of clarity. (a) Averaged \ac{MAE} against the number of decision trees and binning of the calibration grid. (b) Spatial distribution of the \ac{MAE} for different pitches of the calibration grid. (c) Difference of the spatial distribution of the \ac{MAE} for the tested calibration grids. (d) Averaged \ac{MAE} against number of decision trees and number of training events per irradiation position. (e) Averaged \ac{MAE} against number of decision trees and maximum depth. (f) Averaged \ac{MAE} against number of decision trees and learning rate. (g) and (h) Averaged \ac{MAE} against maximum depth and different input sets [raw data (r); raw data and calculated features (r+CF); raw data, calculated features and estimated planar interaction positions (r+CF+pos)]. Ensembles of \num{50} decision trees and \num{1000} decision trees are displayed, respectively.}
\label{fig:general_optimization}
\end{figure*}

We found a maximum depth of \num{10} and a learning rate of \num{0.1} with raw data of \num{5000} training events at \SI{1}{\milli\metre} binning of the calibration grid as a suitable start point. The full optimization process is displayed in Fig. \ref{fig:general_optimization}.

Fig. \subref*{fig:genereal_reference_binning} shows the averaged \ac{MAE} against the number of decision trees and different pitches of the calibration grid. In general, the \ac{MAE} improves with an increasing number of decision trees. Only the model based on \SI{3}{\milli\metre} binning shows slight overfitting effects starting at around \num{400} decision trees. The positioning performance difference is less than \SI{13}{\%} between the \SI{0.25}{\milli\metre} and \SI{3}{\milli\metre} calibration grid evaluated at their respective optimum. For the calibration grids of \SI{0.25}{\milli\metre} and \SI{1}{\milli\metre}, a deviation of the positioning performance of \SI{6}{\%} is found. The course of the spatial distribution of the \ac{MAE} is generally the same for all tested calibration grids [see Fig. \subref*{fig:general_reference_binning_overlay}]: the \ac{MAE} is nearly constant in the central region of the crystal and deteriorates towards the crystal edges. Significant higher order effects such as additional periodicity in the spatial distribution of the \ac{MAE} start to get visible for the \SI{3}{\milli\metre} calibration grid. These effects are not visible for the \SI{1}{\milli\metre} calibration grid; the \ac{MAE} worsens globally. However, the positioning performance difference close to the crystal edges is less dominant [see Fig. \subref*{fig:general_reference_binning_overlay_diff}]. Due to reasons of clarity not shown in Fig. \subref*{fig:genereal_reference_binning} to Fig. \subref*{fig:general_reference_binning_overlay_diff}, further pitches of the calibration grid beside the shown ones were tested - namely \SI{0.5}{\milli\metre}, \SI{0.75}{\milli\metre}, \SI{1.5}{\milli\metre} and \SI{2}{\milli\metre}. All obtained results are in congruence with the plotted pitches of the calibration grid and follow the behavior of the shown curves of averaged and spatial MAE.

Fig. \subref*{fig:general_training_events} displays the averaged \ac{MAE} against the number of decision trees for different numbers of training events per irradiation position. The positioning performance increases if more events are employed for training the \ac{GTB} models. However, only small improvements can be achieved after a sufficient amount of training events is present. Increasing the number of training events per irradiation position from \num{2500} to \num{5000} or \num{10000} improves the positioning performance less than \SI{3.4}{\%} or \SI{6}{\%}, respectively.

The maximum depth strongly influences the positioning performance for ensembles with a low number of decision trees [see Fig. \subref*{fig:general_max_depth}]; increasing the maximum depth improves the averaged \ac{MAE}. Using higher maximum depths than \num{10} show no significant enhancement. The positioning performance difference between the \ac{GTB} models with different maximum depths vanishes for ensembles with a high number of decision trees and converges to an optimum.

Increasing the learning rate, the optimum of the \ac{GTB} models is found for ensembles with less decision trees before overfitting occurs [see Fig. \subref*{fig:general_learning_rate}]. For example, overfitting starts after less than \num{10} decision trees for learning rate \num{0.7} while no overfitting is observed up to \num{1000} decision trees for learning rate \num{0.05}. At the same time, the best possible positioning performance decreases for higher learning rates.

The influence of different input sets at several maximum depths is shown for ensembles of \num{50} and \num{1000} decision trees in Fig. \subref*{fig:general_input_sets_small} and Fig. \subref*{fig:general_input_sets_large}, respectively. For the \ac{GTB} model with \num{50} decision trees, the input sets including the calculated features and planar position improve the averaged \ac{MAE} for all tested maximum depths. The difference between the \ac{MAE} gets smaller increasing the maximum depth. The maximum positioning performance gain by additionally adding the estimated planar interaction positions is smaller than \SI{0.7}{\%} for every maximum depth. The behavior regarding the benefit of adding calculated features and estimated planar interaction positions is similar for \ac{GTB} models with \num{1000} decision trees. In contrast to the \ac{GTB} models with fewer decision trees, the averaged \ac{MAE} slightly deteriorates for higher maximum depths as the models are already in the overfitting regime.

\subsubsection{High-Performance Optimization}

The averaged performance parameters of the best found \ac{GTB} models for all three input sets are displayed in Tab. \ref{tab:overview_results}. A \ac{GTB} model of maximum depth \num{8}, learning rate \num{0.05} and around \num{1550} decision trees performed best for raw data as input. The best models with inputs containing the calculated features for both excluding and including the planar interaction position were found at maximum depth \num{6}, learning rate \num{0.05} and around \num{1000} decision trees. These correspond to memory requirements of \SI{6.7}{\mega\byte} and \SI{1.1}{\mega\byte} according to equation (1), respectively. Furthermore, adding calculated features improves the positioning performance by \SIrange{1.6}{8}{\%} compared to raw data as input while $r_{50}$ and \ac{SR} are affected most. No significant deviation of the positioning performance is induced by adding the planar interaction position.

Fig. \ref{fig:result_IR_GTB} shows the spatial distribution for bias vector, \ac{MAE} and \ac{SR} for raw data and calculated features as input set. The bias vector points towards the crystal center close to both edges. In the region ranging from \SIrange{2}{10}{\milli\metre}, a bias vector magnitude below \SI{1}{\milli\metre} is observed. The \ac{MAE} deteriorates towards the crystal edges and is nearly constant at around \SI{1.2}{\milli\metre} for \ac{DOI} positions between \SIrange{2}{10}{\milli\metre}. The \ac{SR} is nearly constant over the whole crystal and shows a maximum drop of approx. \SI{24.7}{\%} close to \SIrange{6}{8}{\milli\metre}.

\subsubsection{Memory-Requirement Optimization}

\begin{figure}[!t]
\center
\includegraphics[width=3.2in]{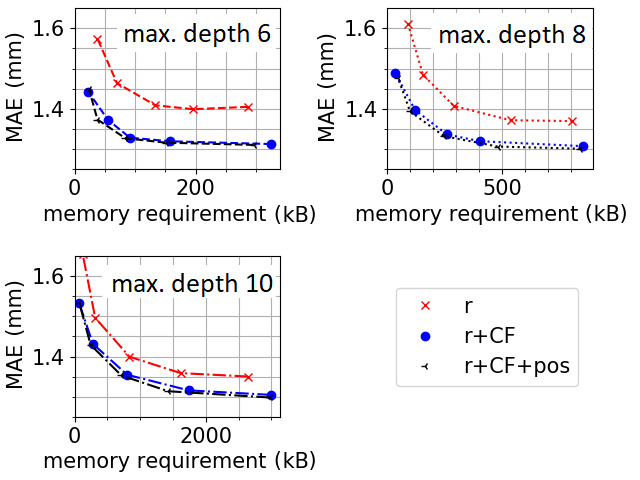}
\caption{Memory-requirement optimization: \ac{MAE} against memory requirement for all three input sets [raw data (r); raw data and calculated features (r+CF); raw data, calculated features and estimated planar interaction positions (r+CF+pos)]. For reasons of clarity, only selected maximum depths are shown. Adding the CF to the input of the \ac{GTB} models improves the positioning performance and allows GTB models with lower memory requirements.}
\label{fig:memory-optimization}
\end{figure}

Fig. \ref{fig:memory-optimization} shows the averaged \ac{MAE} against the memory requirement of \ac{GTB} models. The number of decision trees was determined by the convergence criterion for all three input sets. The lines represent models with the maximum depth denoted next to them while every marker indicates a tested learning rate. The plot allows choosing the best performing \ac{GTB} models for given restrictions on the memory. High learning rates generally led to models with low memory requirements and vice versa. Adding calculated features to the input is beneficial for the positioning performance for models of both low and high memory requirement. The positioning performance improves from \SIrange{0.1}{0.9}{\%} comparing the corresponding \ac{GTB} models with and without estimated planar interaction position of same maximum depth and learning rate.

\section{Discussion}

The photon densities in the employed monolithic crystal are too low to generate a significant amount of events with all \num{16} hits present. This emphasizes the need for calibration and positioning algorithms able to deal with missing hit information. Although the presented results are obtained with the DPC photosensor, both \ac{DOI} estimation methods can be applied to all kinds of photosensors.

The \ac{SO} model can be successfully employed as a \ac{DOI} estimator. In contrast to lookup table-approaches as employed in \cite{vanDam.2011_practicalMethod, Borghi2016, Borghi2016a}, the \ac{IR} model allows a continuous \ac{DOI} estimation. In general, no huge influence on the positioning performance was observed for varying the segmentation of the scintillator because the main differences of the photon detection efficiency between single DPCs are canceled out by the normalization of the \ac{DOI} observable. Here, \ac{SO} models with a segmentation matching the pixel pitch performed best. For more segments, the number of training events per segment decreases below \num{150} per irradiation position which increases the uncertainty of the \ac{IR}. Thus, a finer segmentation may lead to an improved performance in case more training events are available. The positioning performance deteriorates for all performance parameters in the central DOI region (see Fig. \ref{fig:result_IR_GTB}). We assume that the variation of the light distribution is lower in this area reducing the specificity of the used \ac{DOI} observable. A similar course of the positioning performance employing the same \ac{DOI} observable is reported for a \SI{22}{\milli\metre} thick monolithic scintillator \cite{Borghi2016}. The deterioration in the central \ac{DOI} region is significantly reduced if a dual-sided read-out is utilized \cite{Borghi2016a}. In contrast to this work, the calculation of the \ac{DOI} observable in \cite{Borghi2016} is based on fully read out events. The authors presented an adpatation and performance of their method for up to \num{4} missing DPCs. Thus, the observed deterioration could be intensified as the \ac{DOI} observable needs to be limited to one DPC to handle events with missing hits. Van Dam et al. studied a monolithic scintillator of identical height as used in this study testing multiple \ac{DOI} observables \cite{vanDam.2011_practicalMethod}. For all \ac{DOI} observables, a \ac{SR} ranging from \SIrange{1}{5}{\milli\metre} FWHM is reported while a \ac{SR} better than \SI{2}{\milli\metre} FWHM is only shown within \SI{2}{\milli\metre} distance to the photosensor. As far as the \ac{SR} alone allows a comparison, the presented \ac{SO} approach performs better than the presented estimation by van Dam et al. and will be used as one benchmark for the \ac{GTB} models. 

\ac{GTB} models were successfully adapted to the \ac{DOI} estimation problem. The general optimization process shows the influence of the hyperparameters and represents a protocol applicable to further scintillator geometries as well. \ac{GTB} models show well-working regression capabilities allowing to significantly reduce the number of irradiation positions required for training [see Fig. \subref*{fig:genereal_reference_binning} to Fig. \subref*{fig:general_reference_binning_overlay_diff}]. Thus, we were able to reduce the binning of the calibration grid to \SI{1}{\milli\metre} corresponding to \num{12} irradiation positions for a full calibration. Furthermore, \ac{GTB} creates \ac{DOI} positioning models with a comparably small amount of training data in the order of a few thousand events per irradiation position [see \subref*{fig:general_training_events}]. Based on these results, we chose \num{5000} events per irradiation position for all trained models including the high-performance optimization. The selected calibration grid and number of training events per irradiation position lead to a calibration time of less than \SI{30}{\min} for the side irradiation a single detector block. This calibration time seems practical for calibrating a large number of detectors as well. If calibration time is a critical issue, the \ac{GTB} models allow a further reduction of irradiation positions and training events without compromising much on the positioning performance.

After a sufficient maximum depth is reached, a further increase of the maximum depth did not yield to a significantly better positioning performance [see Fig. \subref*{fig:general_max_depth}]. In case a single decision tree corrected the given input during the training process with a smaller depth than the allowed maximum depth, all further added nodes are based on statistical fluctuations found in the training data. As these additional nodes do not contribute to the general predictivity of the \ac{GTB} model, the positioning performance is not improved.

Maximum depth and learning rate are the most important hyperparameters to tune both the positioning performance and memory requirement of the \ac{GTB} models. As the memory requirement is $\mathcal{O}(2^d)$ with $d$ the maximum depth, this is the most important hyperparameter for an FPGA implementation. High learning rates allow a significant reduction of the number of decision trees at the cost of a decreased positioning performance [see Fig. \subref*{fig:general_learning_rate}].

Adding calculated features to the input features improves the positioning performance for \ac{GTB} models with both small and high number of decision trees [see Fig. \subref*{fig:general_input_sets_small} and Fig. \subref*{fig:general_input_sets_large}]. The physically motivated features such as the defined \ac{DOI} observable are easily interpretable and allow faster learning. Also, geometrical information as encoded in the projections has a high information content which is not directly accessible by the raw data. This also leads to less complex \ac{GTB} models with smaller memory requirement as shown for the high-performance optimization. Due to differences in the light response of the detector, the \ac{GTB} models need to distinguish between different planar interaction positions. Thus, it is of interest if the \ac{GTB} models benefit from the planar interaction position concerning a better positioning performance or reduction of required memory. As a consequence, adding the planar interaction position would require a sequential positioning process in a system architecture. As shown in Fig. \subref*{fig:general_input_sets_small} and Fig. \subref*{fig:general_input_sets_large}, only a small positioning performance difference of less than \SI{0.7}{\%} can be observed when adding the planar interaction position to the calculated features. The same observation holds true for the high-performance and memory-requirement optimization (see Tab. \ref{tab:overview_results} and Fig. \ref{fig:memory-optimization}). The estimated planar interaction position does not provide a significant higher information content than the already added calculated features such as center of gravity.

Both presented optimization scenarios focus on different use-cases. In case the calculated features are added to the input set, the high-performance optimization leads to \ac{GTB} models of around \SI{1}{\mega\byte} memory requirement which are easy to handle for modern computers. However, the memory requirement is the limiting factor for currently available FPGA in case no external memory is added. The memory-requirement optimization allows selecting the best-performing model for given resources. For example, the memory requirement can be reduced down to \SI{91}{\kilo\byte} with a reduction of the \ac{MAE} of less than \SI{3}{\%} compared to the high-performance model requiring \SI{1}{\mega\byte}. This memory requirement can be easily handled by currently available FPGA families.

Comparing the \ac{SO} models and high-performance \ac{GTB} models, the \ac{GTB} models have a better positioning performance of up to \SI{50}{\%} for the averaged performance parameters (see Tab. \ref{tab:overview_results}). Particularly, the bias vector and bias-sensitive performance parameters are improved; the averaged \ac{SR} shows no significant difference. However, the differences get clearly visible comparing the spatial distribution of the performance parameters (see Fig. \ref{fig:result_IR_GTB}): In contrast to the \ac{SO} model, the \ac{GTB} models provide a nearly uniform positioning performance over the whole crystal depth. For example, the \ac{SR} ranges between \SIrange{1.9}{2.5}{\milli\metre} FWHM for \ac{GTB} models and between \SIrange{1.0}{5.3}{\milli\metre} for \ac{SO} models. The bias-sensitive performance parameters such as \ac{MAE} [see Fig. \subref*{fig:result_IR_GTB_MAE}] stays constant over the whole crystal except close to the edges for \ac{GTB} models. As a main difference between \ac{SO} and \ac{GTB}, the \ac{GTB} algorithm has access and utilizes all available information of the raw data and caluclated features. To our best knowledge, the \ac{GTB} models achieve the best reported positioning performance for \ac{DOI} estimation with respect to averaged performance parameters and uniformity for the tested scintillator geometry. For example, Li et al. achieve an averaged \ac{SR} of \SI{2.6}{\milli\metre} FWHM (corrected for finite beam width) using a parametric fitting model for a \SI{10}{\milli\metre} thick scintillator \cite{Li2008}. As already discussed, van Dam et al. presented a \ac{SR} ranging from \SIrange{1}{5}{\milli\metre} not providing an uniform positioning performance \cite{vanDam.2011_practicalMethod}. Comparable to our results, Wang et al. demonstrated an averaged \ac{SR} of around \SI{2}{\milli\metre} employing a neural network \cite{Wang2013}. The reported \ac{SR} deteriorated after applying a correction for bias effects. In contrast to the \ac{GTB} method, the monolithic crystal was segmented into \num{35} cuboids with their own neural network \ac{DOI} estimation models. Thus, the planar interaction position needs to be estimated before the \ac{DOI} position.

\section{Conclusion}

We presented two \ac{DOI} positioning methods based on a side irradiation conducted with a fan beam collimator. Both \ac{SO} and \ac{GTB} models handle missing hit information. Due to the regression capabilities of the \ac{GTB} models, the required time for the side irradiation can be reduced to less than \SI{30}{\min} without compromising too much on the positioning performance. The \ac{GTB} model is able to employ the raw photon counts as input set. Adding physically motivated features to the raw data improves the positioning performance and allows models with smaller memory requirements. For \ac{DOI} estimation, the \ac{GTB} models do not require information of the planar interaction position. Thus, a full parallelization of both planar and \ac{DOI} positioning is feasible on system level. The developed memory optimization process allows training \ac{GTB} models suitable for future FPGA implementation of the algorithm. In contrast to the \ac{SO} models and other methods presented in literature, \ac{GTB} models provide a nearly uniform positioning performance over the whole crystal depth. We achieved an averaged MAE of \SI{1.28}{\milli\metre} and SR of \SI{2.12}{\milli\metre} FWHM for the \SI{12}{\milli\metre}-high crystal, respectively. Future research will translate and evaluate the \ac{GTB} algorithm to monolithic scintillators of higher depth. Furthermore, alternatives replacing the side irradiation by means of optical simulation to create the training data are under investigation.

\section*{Acknowledgment}

The authors want to acknowledge the open-source software packages NumPy \cite{VanderWalt2011}, Matplotlib \cite{Hunter2007} and ROOT \cite{Brun1997} heavily used in the analysis of the presented data.

\renewcommand*{\bibfont}{\footnotesize}
\printbibliography[heading=bibnumbered]

\begin{acronym}
\acro{CF}{calculated features}
\acro{COG}{center of gravity}
\acro{DOI}{depth of interaction}
\acro{GTB}{gradient tree boosting}
\acro{IR}{Isotonic Regression}
\acro{kNN}{$k$ nearest neighbors}
\acro{LSF}{line spread function}
\acro{MAE}{mean absolute error}
\acro{ML}{maximum likelihood}
\acro{SO}{single observable}
\acro{SR}{Spatial Resolution}
\acro{SPAD}{single photon avalanche diode}
\acro{PCA}{principal component analysis}
\acro{PDPC}{Philips Digital Photon Counting}
\acro{PSF}{point spread function}
\end{acronym}

\end{document}